\documentclass[conference]{IEEEtran}
\IEEEoverridecommandlockouts

\usepackage{cite}
\usepackage{amsmath,amssymb,amsfonts}
\usepackage{algorithmic}
\usepackage{microtype}
\usepackage{graphicx}
\graphicspath{ {images/} }
\usepackage{booktabs} % for professional tables
\usepackage{graphicx}
\usepackage{textcomp}
\usepackage{subcaption}
\usepackage{xcolor}
\def\BibTeX{{\rm B\kern-.05em{\sc i\kern-.025em b}\kern-.08em
    T\kern-.1667em\lower.7ex\hbox{E}\kern-.125emX}}

\begin{document}

\title{Personalized action suggestions in low-code automation platforms}
\makeatletter
\newcommand{\linebreakand}{%
  \end{@IEEEauthorhalign}
  \hfill\mbox{}\par
  \mbox{}\hfill\begin{@IEEEauthorhalign}
}
\makeatother
\author{
\IEEEauthorblockN{Saksham Gupta}
\IEEEauthorblockA{\textit{PROSE} \\
\textit{Microsoft}\\
Delhi, India \\
t-saksgupta@microsoft.com}
\and
\IEEEauthorblockN{ Gust Verbruggen*}
\IEEEauthorblockA{\textit{PROSE} \\
\textit{Microsoft}\\
Keerbergen, Belgium\\
gverbruggen@microsoft.com}
\thanks{* Listed in alphabetical order}
\and
\IEEEauthorblockN{ Mukul Singh*}
\IEEEauthorblockA{\textit{PROSE} \\
\textit{Microsoft}\\
Delhi, India\\
singhmukul@microsoft.com}
\linebreakand
\IEEEauthorblockN{Sumit Gulwani*}
\IEEEauthorblockA{\textit{PROSE} \\
\textit{Microsoft}\\
Redmond, USA\\
sumitg@microsoft.com}
\and
\IEEEauthorblockN{Vu Le*}
\IEEEauthorblockA{\textit{PROSE} \\
\textit{Microsoft}\\
Redmond, USA\\
levu@microsoft.com}
}

\maketitle

\begin{abstract}
Automation platforms aim to automate repetitive tasks using workflows, which start with a trigger and then perform a series of actions.

However, with many possible actions, the user has to search for the desired action at each step, which hinders the speed of flow development.
We propose a personalized transformer model that recommends the next item at each step.
This personalization is learned end-to-end from user statistics that are available at inference time.
We evaluated our model on workflows from Power Automate users and show that personalization improves top-1 accuracy by 22\%.
For new users, our model performs similar to a model trained without personalization.

\end{abstract}


\section{Introduction}

Workflow automation is a big problem today, but only a fraction of people can write code.
Platforms like Zapier, IFTTT and Microsoft Power Automate allow users to build automated workflows between different applications and services---without a single line of code.
A workflow (or flow) consists of a \emph{trigger} that initiates the flow and the \emph{actions} to be performed.
For example, when an email arrives (trigger) we can store any attachment to a cloud storage provider (first action) and log the email subject in a spreadsheet (second action).

With thousands of actions across different vendors to choose from, even low-code environments can be tedious to use.
Figure~\ref{fig:actionselector} shows the interface for selecting an action in Power Automate.
As flows get longer, the time spent on selecting actions increases.
Modern code development environments boost the productivity of programmers by recommending relevant functions or even whole lines of code \cite{svyatkovskiy2020intellicode}.

\begin{figure}[htb]
\centering
\includegraphics[width=0.5\columnwidth]{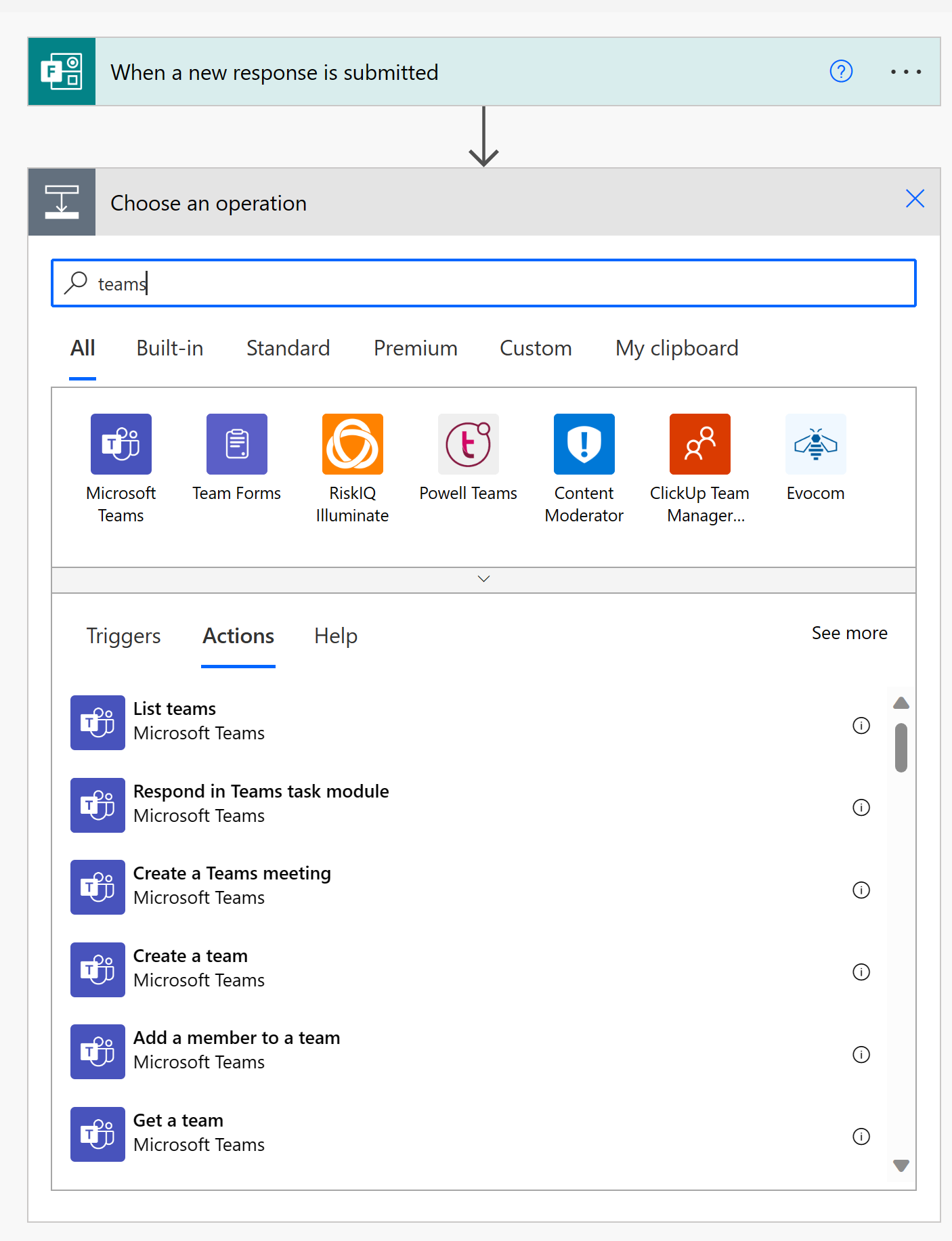}
\caption{Action selection interface in the Microsoft Power Automate platform. For each action in a workflow, the user must manually navigate to the desired action.}
\label{fig:actionselector}
\end{figure}

In this paper, we tackle the problem of recommending relevant actions to the user.
Our approach is inspired by recent advances in code completion \cite{codex,codefill} and uses a transformer model to predict the distribution over likely next actions.

A key difference with code completion is that personalization depends more on the user and less on the context.
Whereas code completion can use context (imported packages or function definitions) to determine likely functions, suggested actions depend more on user preferences, for example, vendors or actions that they have used in different flows.
To make personalized suggestions, we embed usage statistics into a personalization vector and learn this embedding end-to-end by feeding it as a token to the decoder model.
New users will not have a personalization vector, and we ensure consistent suggestions by sometimes providing the model with an empty personalization vector.

We focus on the Power Automate platform and use 674K real flows to evaluate our model.
We show that learning personalization vectors improve performance by 14\% over personalization during inference (based on vendors) and by 22\% over no personalization.
Additionally, we show how the performance on new users remains consistent.

In summary, we make the following contributions:
\begin{enumerate}
    \item We introduce a personalized decoder-only transformer that learns user profiles end-to-end.
    \item We ensure that suggestions for new users are on par with a non-personalized model by varying the personalization rate during training.
    \item We train and evaluate our model on 600K and 74K flows respectively, showing that personalization improves performance by 22\% in top-1 recommendations, showing that personalization does not affect suggestions for new users, and that learning personalization vectors yields better suggestions than inference personalization. 
\end{enumerate}

\section{Related Work}

Suggesting actions in automation platforms is a novel problem, but this is closely related to the task of predicting words in natural language or making suggestions in code.

Early language modeling uses $n$-grams to predict the next token.
Combining different $n$ though complicated backoff schemes was shown to significantly boost performance \cite{rosenfeld2000two}.
With enough data, a backoff schedule called ``stupid backoff'' \cite{stupidbackoff} which simply uses the largest $n$ for which a prefix exists, was shown to work very well.
The downside of $n$-gram models is that they give the same attention to each token (word or action) in the input context and completely ignore inter-token dependencies which are often critical for understanding tasks.

Language models based on transformer architectures \cite{vaswani2017attention} have greatly improved performance in a wide spectrum of language understanding and generation tasks by incorporating inter-token dependencies via attention.

It was shown that these models outperform humans at predicting tokens in a causal setting \cite{transformersbetter}.
Decoder-only architectures are particularly well suited for the task of predicting the next token given a history of tokens \cite{formaltransformers}.
Based on their success, we use a decoder-only language model \cite{gpt} as the base architecture for our system.

Similar to suggesting an action in a workflow, code recommendation is an active research area that has been integrated in many commercial products.
For example, GitHub Copilot \cite{codex} uses a decoder trained on code to suggest whole blocks of code.
IntelliCode in Visual Studio uses a similar model to suggest line completions \cite{svyatkovskiy2020intellicode} and is small enough to run on the client.
These models offer personalization through prompting (Copilot) or require fine-tuning on the repositories for large organizations (IntelliCode).

We take inspiration from SSE-PT \cite{ssept}, which introduces personalization using transformers.
In SSE-PT, user embeddings are appended to all token embeddings of the input sequence.
To avoid overfitting, they rely on regularization by applying Shared Stochastic Embeddings \cite{sse}.
In contrast, we only prepend the user embeddings at the start of our input sequence as a single token and do not require any regularization.
By keeping user information at the token level, the model learns to selectively attend to personalization tokens based on the input.

\section{Method}

We use a decoder-only architecture \cite{gpt} as predicting the next action closely aligns with the auto-regressive pre-training objective.
A personalization vector is passed to the decoder to make personalized suggestions.
An overview of our architecture is shown in Figure~\ref{fig:generalmodel}.
The following two sections describe how flows are tokenized, and how this decoder model is adapted for learning to make personalized suggestions end-to-end.

\begin{figure}[tbh]
\centering
\includegraphics[width=0.8\columnwidth]{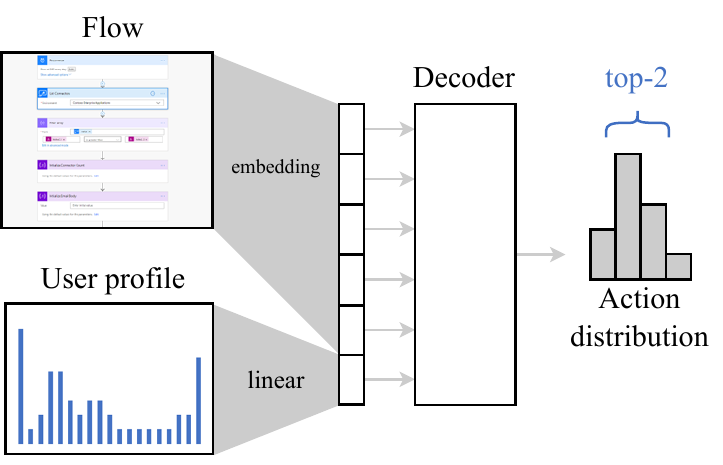}
\caption{Architecture diagram for the proposed \textit{Personalized Decoder} model. The model leverages both the current flow and user history to predict the next action in the flow.}
\label{fig:generalmodel}
\end{figure}

\subsection{Tokenizing Flows}

A \emph{flow} is a rooted directed acyclic graph where the root is the trigger and all other nodes are actions.
There are different types of actions, the most common of which are control flow statements (which cause the graph to not always be linear) and API actions.
API actions consist of a connection and an operation, for example, Outlook (connection) and SendEmail (operation).
An example flow is shown in Figure~\ref{fig:exampleflow}.

\begin{figure}[tbh]
\centering
\includegraphics[width=0.6\columnwidth]{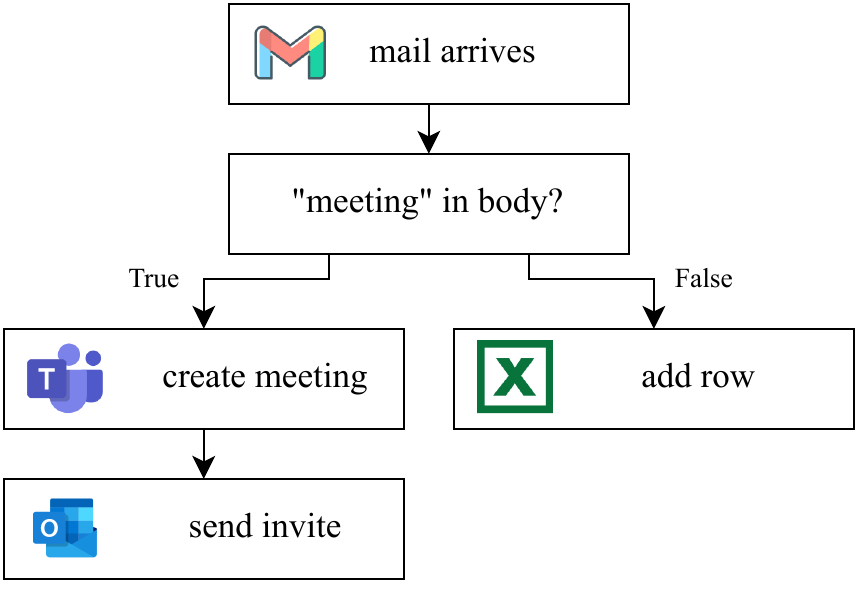}
\caption{Example flow that, after a mail arrives, either creates a meeting and sends an invite, or adds a row to an Excel file.}
\label{fig:exampleflow}
\end{figure}

The context when making a prediction for an action are its parent actions.
In other words, when making predictions in one branch, we do not look at actions from other branches.
This sequence of actions is called the \emph{prefix}.

As opposed to code and language, the vocabulary for actions is closed---every flow is a combination of the same 1423 actions. 
We therefore do not use subword tokenization \cite{sennrich-etal-2016-neural} and keep every action as a separate token.
During development, we found that even only using separate tokens for connection and operation (1) yields token predictions that are hard to aggregate into one action, and (2) is significantly slower due to requiring many individual token predictions.

\subsection{Personalization} 

We map user information to a personalization vector and prepend this to the embedded flow.
The user information we consider is a distribution of actions that the user has used in previous flows---other information like demographic features can be added, either to this vector or as a new vector.

The personalization vector is trained end-to-end with the model.

To evaluate whether these profile vectors retain some of the information about action usage, we reduce them to two dimensions through PCA \cite{pca} and plot them.
In Figure~\ref{fig:usermap}, we color each point based on the proportion of actions related to Microsoft products, as these actions are the most common by far.
The clear gradient indicates that profile vectors exploit action counts and learn which actions are related.
In Figure~\ref{fig:usermap_twitter}, only distinguish (92) users that have used any Twitter action and a randomly sampled set of (460) other users.
Some clustering is clearly present, indicating that rare connections are also linked in the profile vectors.

\begin{figure}[tbh]
\centering
\includegraphics[width=0.99\columnwidth]{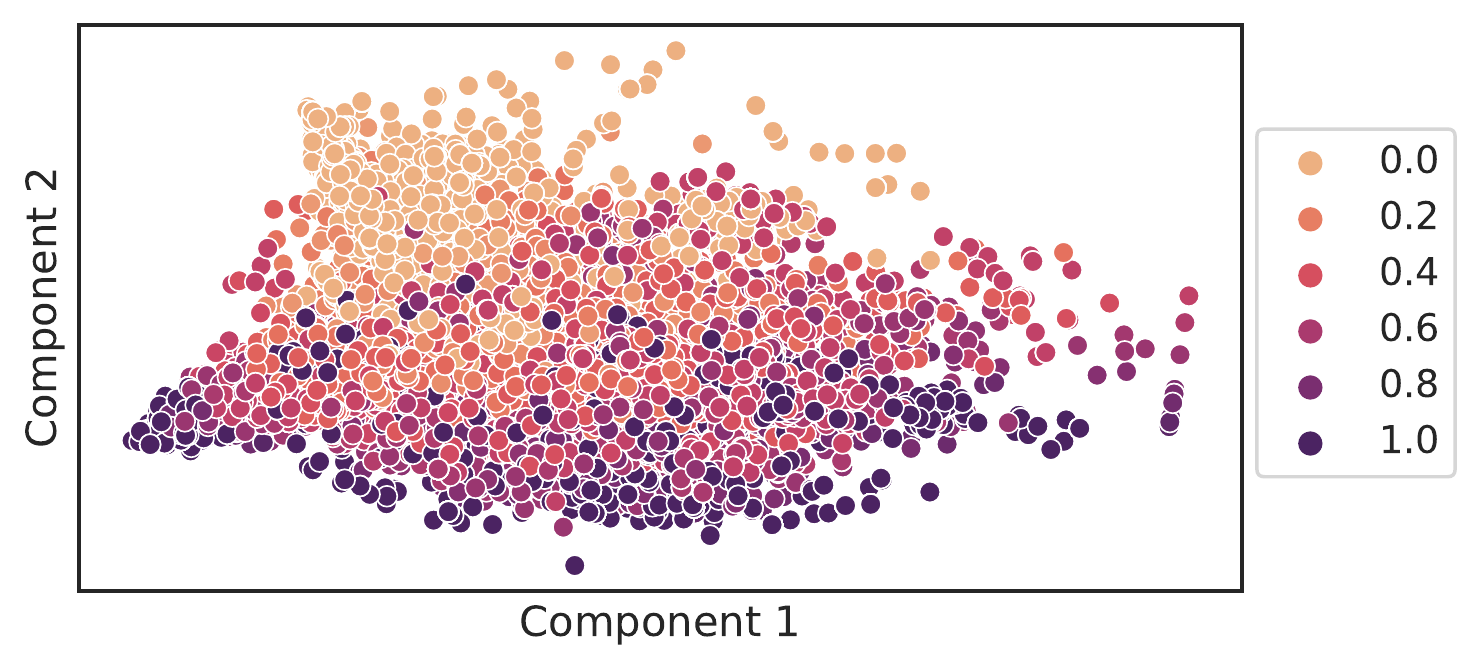}
\caption{User Embeddings for users with different percentage of \textit{Microsoft} based actions in their history. 0 means no \textit{Microsoft} action while 1 means only \textit{Microsoft} actions in history. There is a clear separation between users based on their action history showing that the models captures these usage patterns.}

\label{fig:usermap}
\end{figure}
\begin{figure}[tbh]
\centering
\includegraphics[width=0.85\columnwidth]{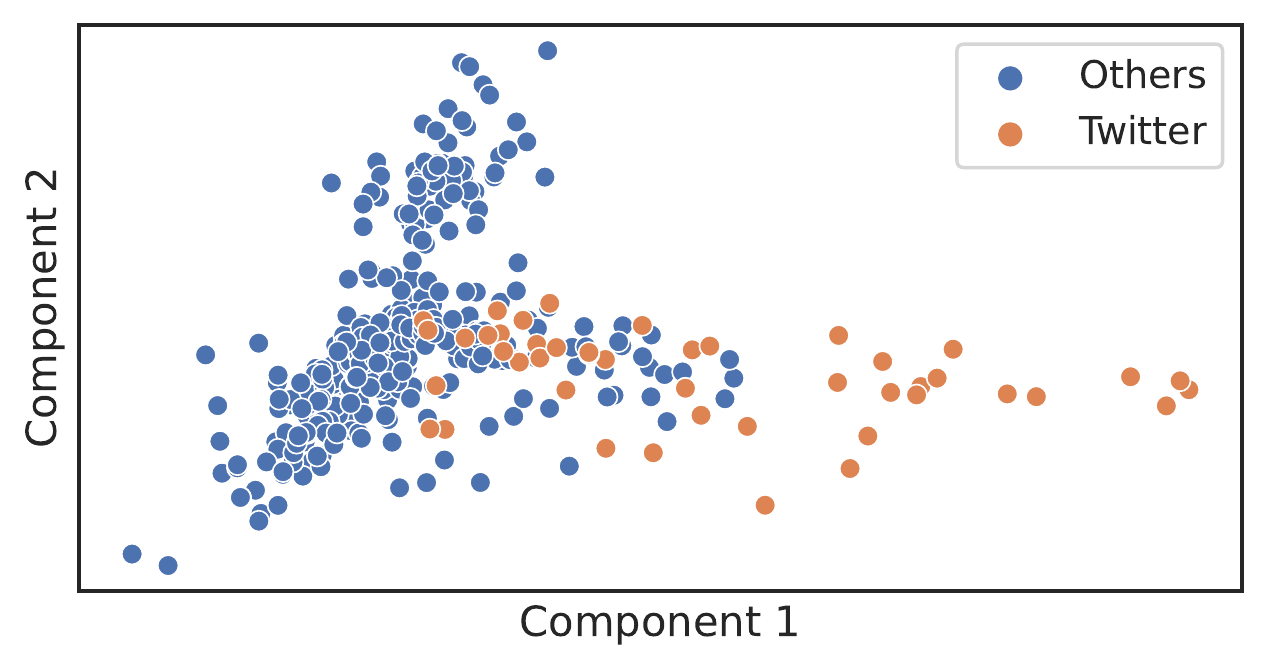}

\caption{User Embeddings for users with Twitter based actions compared against other users. We plot the first 2 principal components. The separation between Twitter and other users show that the user embeddings capture platform information.}
\label{fig:usermap_twitter}
\end{figure}

\section{Evaluation and Results}

We perform experiments to answer the following research questions:
\begin{enumerate}
    \item[\bfseries Q1.] Does personalization help in recommending actions?
    \item[\bfseries Q2.] Does the personalized model make good predictions for new users?
    \item[\bfseries Q3.] Case study: can we limit the number of recommendations based on the predicted probabilities? 
\end{enumerate}

\subsection{Evaluation Setup}

Here we describe the setup used to evaluate our model and the custom baselines we compare our model against.

\subsubsection{Data} We divide our 674K flows by user to ensure that no user has a flow in both train (600K) and test (74K) sets.

\subsubsection{Evaluation}

The model returns a distribution over actions, which we consider as a ranking.
As common in both code generation and recommendation methods, we evaluate whether the actual action is amongst the top-$k$ ranked ones or not.
This neatly aligns with the possibility of multiple actions following the same prefix.

\subsubsection{Baselines}

Besides our transformer model, we use two other approaches to evaluate the problem and our solution.

\begin{itemize}
    \item We train an \textbf{n-gram} with stupid backoff, which computes the probability $p(a_{i+1} \mid a_{i-n} \ldots a_{i})$ of an action based on how often $a_{i+1}$ follows the prefix $a_{i-n} \ldots a_{i}$ and falls back to $n-1$ if that prefix does not exist. This (very) roughly corresponds to a decoder that is not able to give more weight to specific actions.
    \item Given a prefix, the top-$k$ \textbf{theoretical maximum} is the proportion of next actions that are in the top-$k$ most common ones in the test set. For example, a prefix with continuations [a, b, b, c, c, c] has a top-1 theoretical maximum of 50\%. Essentially, this corresponds to a model that is perfect with respect to the testing data.
    \item We train a \textbf{simple} decoder model without personalization.
    \item We perform personalization at \textbf{inference} time with the simple model, either by only allowing API actions of \textbf{connections} that the user has previously used, or by weighing the predicted probabilities by how often the user has used this \textbf{action} in the past.
\end{itemize}

\subsubsection{Hyperparameters} Our decoder is lightweight, with two layers and an embedding dimension of 256 spread over two attention heads---resulting in 2.3M parameters. The stupid backoff model uses n = 5, which roughly corresponds to the same number of parameters.

\subsection{Personalization (Q1)}

\begin{figure}[tbh]
\centering
\includegraphics[width=0.85\columnwidth]{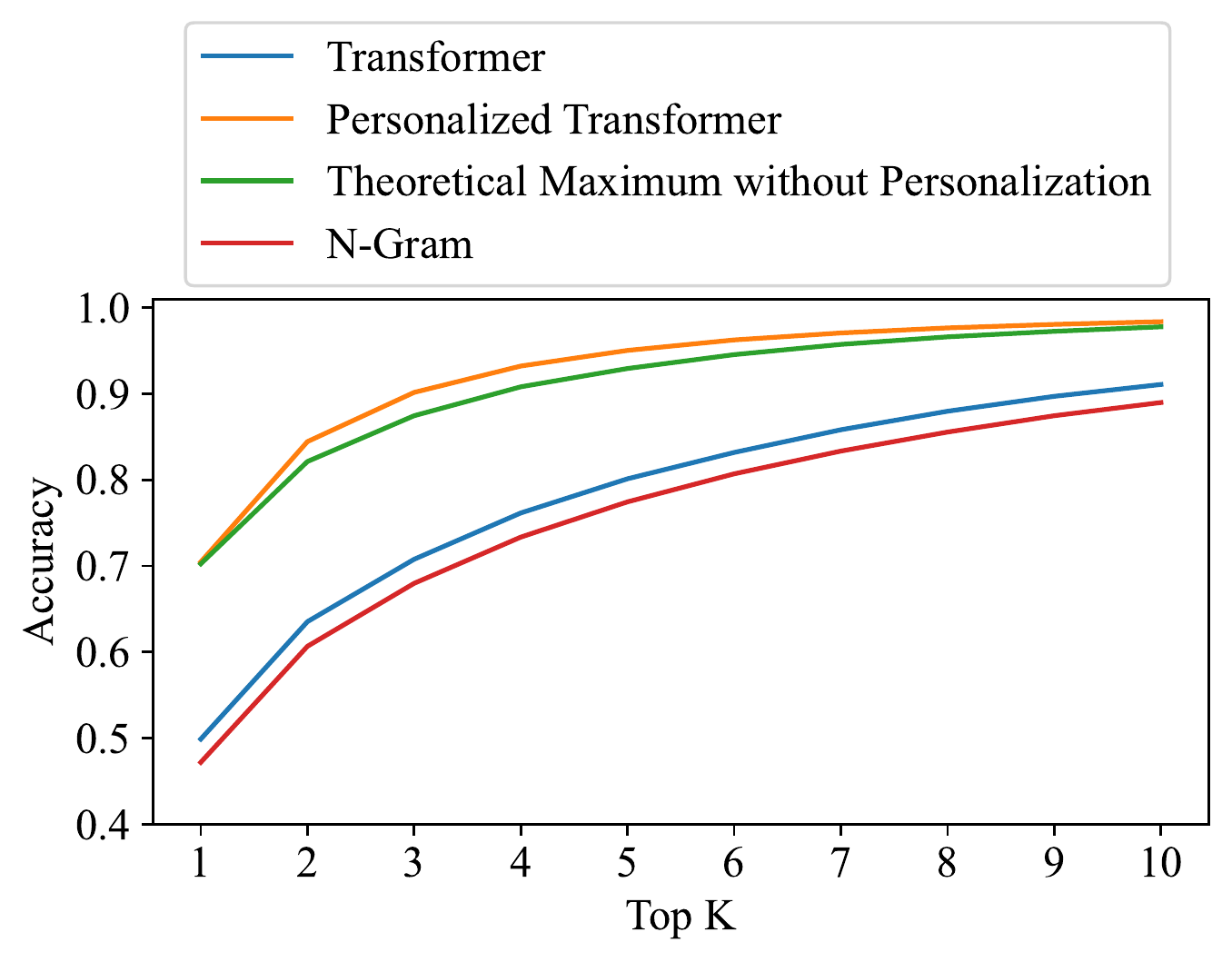}
\caption{Comparing top-$k$ action prediction accuracy for our system against baselines. We plot results for increasing values of $k$. Personalized transformer performs the best and is significantly better than simple transformer model.}
\label{fig:baselinecomparison}
\end{figure}

Figure~\ref{fig:baselinecomparison} shows the top-$k$ performance for increasing $k$.
Our personalized model performs on par or better than the theoretical maximum without personalization.
When making three suggestions, the desired action is recommended 90\% of the time.
When making ten suggestions, this increases to 98\%.
The simple transformer (blue) barely outperforms the n-gram model, indicating that this is a challenging task.

\begin{figure}[tbh]
\centering
\includegraphics[width=0.85\columnwidth]{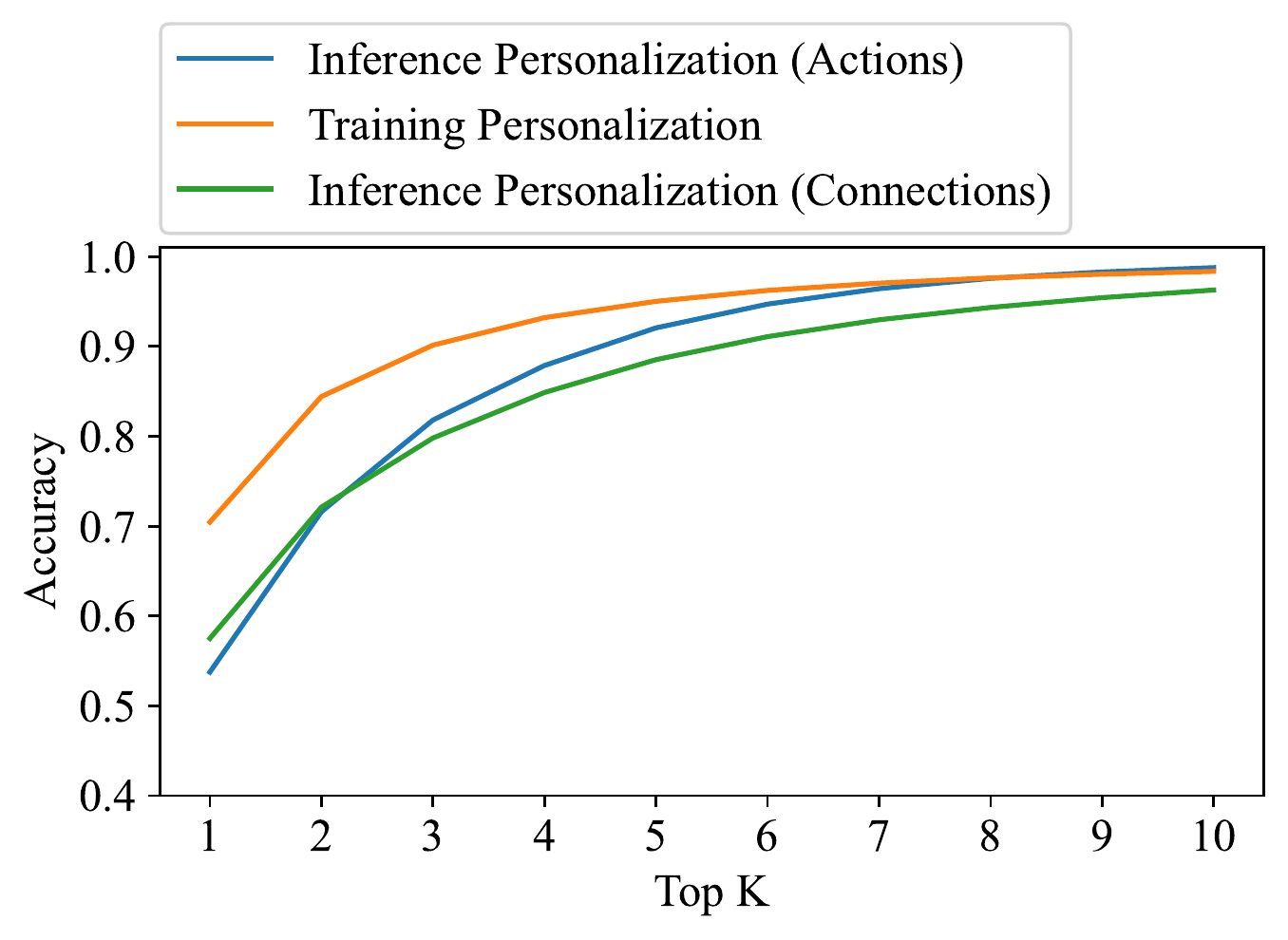}
\caption{Top-$k$ action prediction accuracy for different values of $k$ using different personalization strategies. Adding personalization during training performs the best with significant improvement in top-1 accuracy.}
\label{fig:personalizationcomparison}
\end{figure}

Figure~\ref{fig:personalizationcomparison} shows that personalization at inference time performs significantly worse---about 14\% percentage points for top-1.
Action personalization performs better than only using connections after showing three or more suggestions, indicating that users often reuse actions.
This hypothesis is reinforced by action personalization performing similar to trained personalization model in top-10 accuracy, both containing the desired action 98.5\% of the time.

\subsection{New Users (Q2)}

New users do not have a profile, but we still want to make good recommendations for them.
Figure~\ref{fig:personalizationpercentagecomparison} shows top-1 (green) and top-4 (orange) accuracy on users with (dashed) and without (full) profiles, for varying levels of personalization during training.
For example, with 50\% personalization, for half examples that the model sees the user profile as all zeros.

Full personalization decreases performance on new users by about 18\%.
Without personalization, users with and without profiles get the same performance---there is no leakage between training and testing.
The exact degree of personalization has minimal influence on the results---it was left at 50\% for all other models.

\begin{figure}[tb]
\centering
\includegraphics[width=0.9\columnwidth]{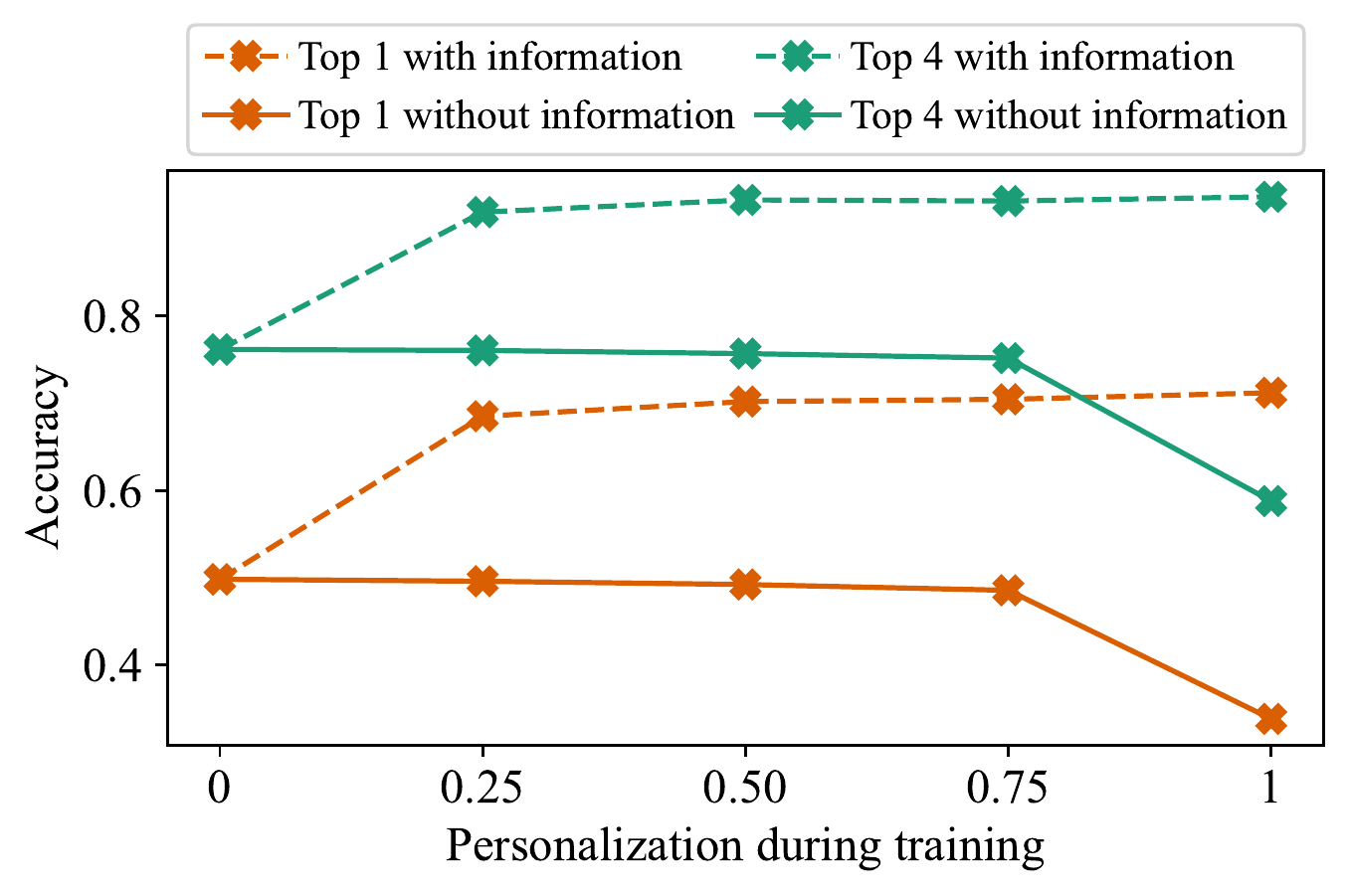}
\caption{\textit{Top-1} and \textit{top-4} prediction accuracy compared against increasing levels of personalization during training for existing and novel users. Adding personalization improves performance for existing users without losing performance for novel users}
\label{fig:personalizationpercentagecomparison}
\end{figure}

\subsection{Limiting Recommendations (Q3)}

To reduce false positives, we want to suppress recommendations when the model is not confident \cite{flashfillpp}.
We provide some preliminary insights into the relation between predicted probabilities and the position of the desired action in Figure~\ref{fig:logprobsk}.
Only making predictions when this probability exceeds some threshold allows us to limit the number of undesired suggestions.

\begin{figure}[tbh]
\centering
\includegraphics[width=0.85\columnwidth]{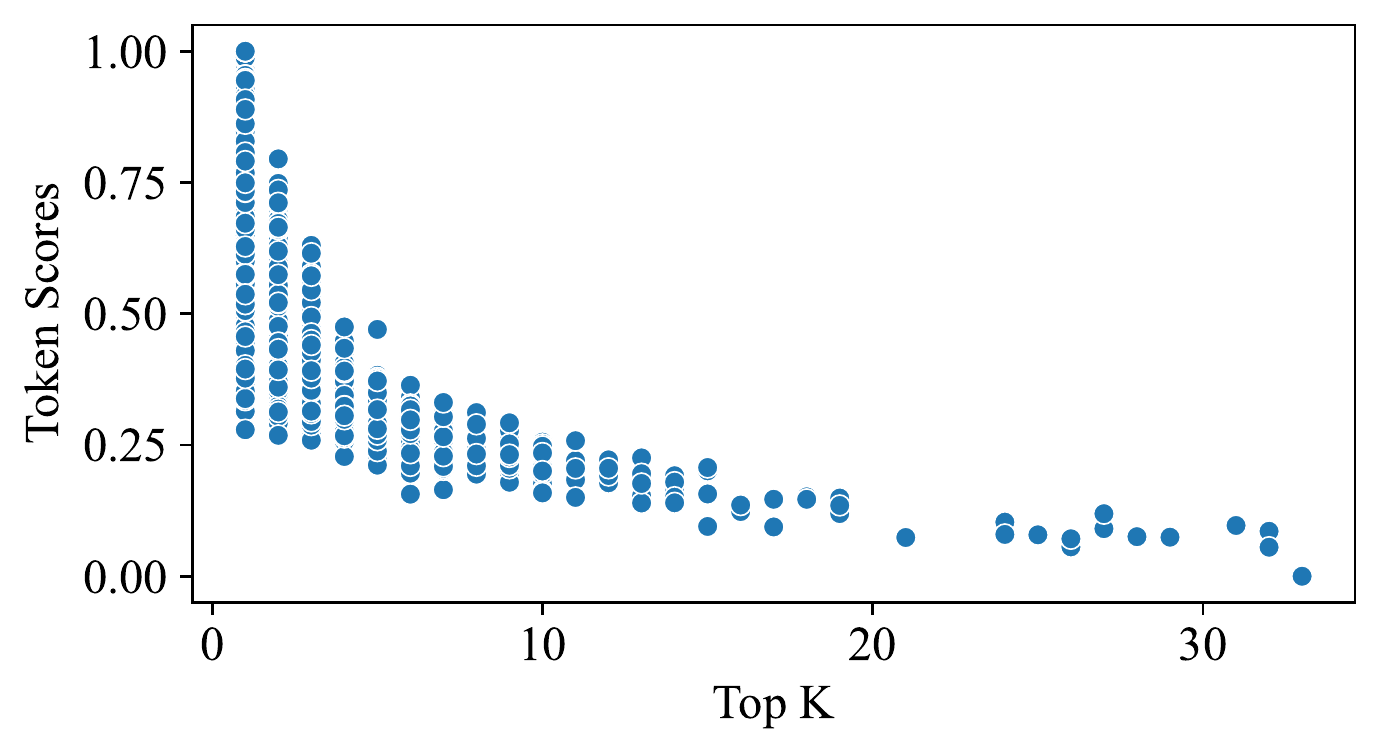}
\caption{Probabilities for tokens being predicted at position K.}
\label{fig:logprobsk}
\end{figure}
 
\section{Conclusion}

We introduced a personalized decoder model for recommending the next best action in automation platforms.
Our model learns to embed action statistics to user personalization vectors end-to-end by considering them as embedded tokens.
Our experiments show that learning personalization vectors significantly improves performance over not making personalized suggestions, or only personalizing predictions during inference. 
Personalization does not cause worse predictions by not always providing user information during training.
In the future, we want to use the predicted action probabilities to suppress recommendations that the model is not certain about.

\bibliographystyle{plain} % We choose the "plain" reference style
\bibliography{references} % Entries are in the refs.bib file

\end{document}